# Experimental probe of weak value amplification and geometric phase through the complex zeroes of the response function

Mandira Pal[1], Sudipta Saha[1], Athira B S[2], Subhasish Dutta Gupta[3] and Nirmalya Ghosh[1,*]

[1]*Department of Physical Sciences,*

*Indian Institute of Science Education and Research (IISER) Kolkata.*

*Mohanpur 741246, India*

[2]*Center of Excellence in Space Sciences India,*

*Indian Institute of Science Education and Research (IISER) Kolkata.*

*Mohanpur 741246, India*

[3]*School of Physics, University of Hyderabad, Hyderabad 500046, India*

*Corresponding authors: nghosh@iiserkol.ac.in*

**Abstract**

The extraordinary concept of weak value amplification has attracted considerable attention for addressing foundational questions in quantum mechanics and for metrological applications in high precision measurement of small physical parameters. Here, we experimentally demonstrate a fundamental relationship between the weak value of an observable and complex zero of the response function of a system by employing weak value amplification of spin Hall shift of a Gaussian light beam. Using this relationship, we show that arbitrarily large weak value amplification far beyond the conventional weak measurement limit can be experimentally obtained from the position of the minima of the pointer intensity profile corresponding to the real part of the complex zero of the response function. The imaginary part of the complex zero, on the other hand, is related to the spatial gradient of geometric phase of light, which in this particular scenario evolves due to the weak interaction and the pre and post selections of the polarization states. These universal relationships between the weak values and the complex zeros of the response function may provide new insights on weak measurements in a wide class of physical systems. Extremely large weak value amplification and extraction of system parameters outside the region of validity of conventional weak measurements may open up a new paradigm of weak measurement enhancing its metrological applications.

Weak measurement, after being discovered by Aharonov, Albert and Vaidman (AAV) in their seminal work [1], remains to be a rather enigmatic and hotly debated topic in physics [2-6]. This measurement process involves preparation of the system in a definite initial state, which due to weak coupling to the observable results in a superposition of "slightly" separated eigenstates. Subsequent post selection in a final state which is nearly orthogonal (small parameter $\epsilon \sim 0$ off from the orthogonal state) to the initial state leads to the outcome, the so-called 'weak value'. Contrary to ideal 'strong' measurement, the weak value can lie outside the eigenvalue spectrum of the observable and can also assume complex values. These extraordinary features of weak measurement have made it an attractive approach in quantum measurements, e.g., for probing quantum paradoxes, for direct measurement of quantum states etc. [6-14]. This concept can be understood using the wave interference phenomena and is therefore equally applicable to interference of quantum matter waves and classical electromagnetic waves. Indeed, the first experimental observation of spin Hall effect of light was achieved using weak measurements in classical optical setting [15]. A plethora of experiments and experimental proposals have followed to exploit the unprecedented ability of weak measurements to faithfully amplify and observe small physical parameters, making it an attractive tool for metrological applications in the optical domain and beyond [16-18]. The applications include detection of ultrasensitive beam deflections [13,15,19], high precision measurements of angular rotation [20], phase shift [21] , temporal shift [22,23] , frequency shift [24] , and so forth.

The weak values are usually extracted from the shift of the pointer profile, which is conventionally taken to be Gaussian, e.g., Gaussian spatial modes of laser beams or Gaussian temporal pulse used in the optical domain [15,22]. From AAV's approximation using Gaussian pointer, there is however, a limit of the minimum value of the small off-set parameter $\epsilon$ that can be used for weak value amplification ($\epsilon_{min}$) [25,26]. Depending upon the shape of the pointer, this limit in turn sets a bound on the maximum achievable weak value amplification ($\sim \frac{1}{\epsilon_{min}}$) [27]. For practical purposes it is therefore important to explore ways to achieve weak value amplification that can exceed this conventional limit. On conceptual ground it is also equally important to understand the weak values through physically meaningful and experimentally accessible properties such as the system response function [28]. Here, we have addressed these two important issues pertaining to weak measurements. Taking an example of weak value amplification of momentum domain spin Hall shift of Gaussian light beam [15], we first establish a relationship between the weak value of the polarization (spin) observable and the complex zero of the spatial response function of the system. The relationship between the weak value and the real part of the complex zero of the response function opens up an avenue to obtain weak value amplification beyond the conventional limit using the position of the minima of the pointer intensity profile. The imaginary part of the complex zero of the response function, on the other hand, is related to the inverse spatial gradient of geometric phase of light, quantification of which also provides direct information on the weak value. We experimentally demonstrate both the concepts by performing weak measurements on Spin Hall shift of Gaussian light beam undergoing partial reflection at dielectric interfaces. The intriguing relationships between the weak values and the complex zeros of the system response function are universal and may provide new insights and understanding on weak measurements in a wide class of physical systems. The experimental demonstration of weak value amplification outside the validity domain of conventional weak measurement may also have significant impact on weak measurement-based metrological applications.

We consider the weak interaction case of spin Hall (or Imbert–Fedorov) shift of fundamental Gaussian beam due to partial reflection at a dielectric interface. The corresponding polarization operator for the spin Hall (SH) shift is defined as [29,30].

$$A^{SH} = \begin{bmatrix} 0 & i\left(1+r_p/r_s\right)\cot\theta_i \\ -i\left(1+r_s/r_p\right)\cot\theta_i & 0 \end{bmatrix} \qquad (1)$$

Here, $\theta_i$ is the angle of incidence and $r_p$, $r_s$ are the angle-dependent amplitude reflection coefficients for $p$ and $s$ linear polarizations, respectively. In order to avoid the interplay of the other variant of the beam shift, namely, the Goos-Hänchen (GH) shift in the weak value amplification, we chose the input or the pre-selected polarization state to be $|\psi_{in}\rangle = |\psi_{pre}\rangle = [1, \ 0]^T$ ($p$ linear polarization), which is the eigen state of the GH shift [31]. The post selections are done at nearly orthogonal linear polarizations $|\psi_{post}\rangle = [\pm\sin(\epsilon), \ \cos(\epsilon)]^T$. Here, $\epsilon$ is a small angle by which the post selected state is away from the orthogonal to the input state. The corresponding expression for the weak value of the SH shift can be written as [29]

$$A_W^{SH} = \frac{\langle\psi_{post}|A^{SH}|\psi_{pre}\rangle}{\langle\psi_{post}|\psi_{pre}\rangle} = \pm iRcot\epsilon \qquad (2)$$

where $R = \left(1 + r_s/r_p\right)\cot\theta_i$. Here, we have used Jones vector $|\psi\rangle$ to represent the polarization state of light and have adopted quantum notations because of its inherent mathematical simplicity of representing weak measurement [29,32]. Note that the eigen states of the SH shift are left/right circular (elliptical) polarizations, which yield real eigenvalues corresponding to shift in the co-ordinate ($y$) space (spatial shift). Weak value amplification as per Eq. 2, on the other hand, yields imaginary weak value representing shift in the conjugate transverse momentum ($p_y$) space (angular shift). This angular shift is detected as a shift of the centroid ($\Delta y$) of the Gaussian beam pointer at the detection plane, which can be related to the weak value as [32]

$$\Delta y = \frac{\lambda}{2\pi}\frac{z}{z_0}\left[Im(A_W^{SH})\right] \qquad (3)$$

Here, $\lambda$ is the wavelength, z is the propagation distance, $z_0$ is the Rayleigh range ($z_0 = \frac{\pi w_0^2}{\lambda}$), and $w_0$ is the beam waist. The above expression for the weak value amplified shift of the Gaussian beam centroid is bounded by a minimum value of the small off-set angle ($\epsilon_{min} \sim \frac{\delta_p}{w_0}$), where $\delta_p = \frac{\lambda}{2\pi}R$ [15]. Beyond this limit ($\epsilon < \epsilon_{min}$), extraction of the weak value from the shift of the beam centroid does not yield the desired amplification as predicted by Eq. 2 (subsequently discussed and experimentally demonstrated) [25].

In order to establish a connection between the weak value of the polarization observable and the zero of the system response function, we use the generalized expression for the reflected field vector from a dielectric interface for an incident Gaussian beam with linear polarization state $[a_p, \; a_s]^T$ [32]

$$\vec{E}_r \propto \exp\left[k\left(iz - \frac{x^2 + y^2}{2\left(z_0 + iz\right)}\right)\right] \begin{pmatrix} r_p\left(1 - i\frac{x}{z_0 + iz}\frac{\partial \ln r_p}{\partial \theta_i}\right) & i\frac{y}{z_0 + iz}\left(r_p + r_s\right)\cot\theta_i \\ -i\frac{y}{z_0 + iz}\left(r_p + r_s\right)\cot\theta_i & r_s\left(1 - i\frac{x}{z_0 + iz}\frac{\partial \ln r_s}{\partial \theta_i}\right) \end{pmatrix} \begin{pmatrix} a_p \\ a_s \end{pmatrix} \quad (4)$$

Here, $x$ is the coordinate in the plane of incidence, while $y$ is perpendicular to this plane. As previously noted, in this case, weak measurement is performed exclusively on transverse (along $y - direction$) Spin Hall shift by nullifying the in-plane (along $x - direction$) GH shift using pre-selection in $p$ linear polarization state. Thus, the reflected field amplitude relevant to the weak value amplification of Spin Hall shift for post selection states $\pm\epsilon$ can be obtained by setting $x = 0$ as

$$E_r^w(y) \propto \exp\left[-\frac{ky^2}{2(z_0+iz)}\right] \times G(y); G(y) = \left[1 \pm i\frac{y}{z_0+iz}R\cot\varepsilon\right] \qquad (5)$$

The $1^{st}$ term can be interpreted as the Gaussian field amplitude of the reflected virtual beam and the $2^{nd}$ term ($G(y)$) encodes information on the change in the field distribution due to the weak interaction and post selection. $G(y)$ may thus be treated as the spatial response function. Note that this space dependent response function for momentum domain Spin Hall shift is in precise analogy with the frequency response of the system corresponding to time delay or frequency shifts of Gaussian temporal pulse [33,34]. The complex root of the response function can be obtained as

$$y_0 = \mp\frac{z-iz_0}{R\cot\varepsilon} \qquad (6a)$$

$$y_0^{real} = \mp\frac{z}{R\cot\varepsilon} = -\frac{z}{\text{Im}\left(A_{SH}^w\right)} \qquad (6b)$$

$$y_0^{imag} = \pm\frac{z_0}{R\cot\varepsilon} = \frac{z_0}{\text{Im}\left(A_{SH}^w\right)} \qquad (6c)$$

Clearly for $\epsilon \neq 0$, the root is complex. The weak measurement is performed around this complex zero or in other words post selections at $\pm\epsilon$ moves the zero from the upper to the lower half complex position ($y$) plane. Complex root also implies that the amplitude (or intensity) of the reflected wave never becomes zero; rather it reaches a finite minimum. This spatial position of this intensity minimum is defined by the real part of the root. As the value for $\epsilon$ is reduced to zero, the response function meets a real zero at $y_0 = 0$ and then the pointer intensity profile becomes double humped Gaussian [25]. With increasing $\epsilon$ within the familiar weak measurement limit for Gaussian pointer ( $\epsilon > \epsilon_{min} \sim \frac{\delta_p}{w_0}$), the intensity distribution turns out to be approximately a Gaussian with its centroid shifted by $\Delta y$ proportional to the weak value ($\propto cot\epsilon \sim \frac{1}{\epsilon}$ in the limit of small $\epsilon$) as per Eq. 3. However, in the limit $\epsilon < \epsilon_{min}$, the weak measurement approximation breaks down and the amplification effect saturates [25,27]. The real part of the complex zero of the response function (Eq. 6b) may then be used to extract the weak value information. This opens up a remarkably simple yet unexplored avenue of obtaining arbitrarily large weak value amplification that may lie far beyond the conventional weak measurement limit ($\epsilon << \epsilon_{min}$) for a given pointer profile.

The information on the imaginary part of the root can be obtained from phase measurements as it is related to the geometric phase of light that evolves during weak measurements. In this particular scenario, the pre ($|\psi_{pre}\rangle = [1, \ 0]^T$) and the post-selected states ($|\psi_{post}\rangle = [\pm\sin(\epsilon), \ \cos(\epsilon)]^T$) are homogeneous states. In contrast, the intermediate state following the weak interaction (reflection) is inhomogeneous (space varying) and its expression can be obtained from Eq. 4 as $|\psi_{int}\rangle = [r_p, \ -i\frac{y}{z+z_0}(r_p + r_s)cot\theta_i]^T$. This sequential evolution of the polarization state during the weak measurement leads to the generation of Pancharatnam-Berry (PB) geometric phase ($\Phi_{PB}$) and its expression can be worked out using Pancharatnam's connection [35,36]. In the limit $z << z_0$, the corresponding expression becomes

$$\Phi_{PB} = Arg\left(\langle\psi_{post}|\psi_{pre}\rangle\langle\psi_{pre}|\psi_{int}\rangle\langle\psi_{int}|\psi_{post}\rangle\right) \approx \pm\frac{yRcot\epsilon}{z_0}$$

$$\frac{\mathrm{d}\Phi_{PB}}{\mathrm{d}y} = \frac{Im(A_W^{SH})}{z_0} = \frac{1}{y_0^{imag}} \tag{7}$$

Apparently, the transverse spatial ($y$) gradient of PB geometric phase leads to a large shift in the transverse momentum ($p_y$) distribution of the Gaussian beam, manifesting as weak value amplified momentum domain beam shift. Importantly, Eq. (6c) and (7) provides an interesting way of quantifying the weak value using interferometric measurement. In what follows, we provide experimental demonstration of the above two concepts.

Note that Eq. 6 and 7 are derived for the specific case of weak measurements on momentum domain Spin Hall shift of Gaussian light beam, where the weak value is purely imaginary. Accordingly, the zero of the response function and the spatial gradient of geometric phase are related to the imaginary weak value of the polarization operator. While the exact algebraic expressions connecting these entities depend upon the specific problem, such relationships are universal and may be formulated for a wide class of weak measurement systems with appropriate consideration of the relevant parameter space, observables, choice of the pointer (Gaussian or non-Gaussian) and the general complex nature of weak values [1,18,20,37].

For probing the real part of the complex zero of the response function, we employed weak measurement on Spin Hall shift of Gaussian light beam undergoing partial reflection at air-glass interface (**Figure 1a**). The fundamental Gaussian mode of 632.8 nm line of a He-Ne laser was used to seed the system. A rotatable linear polarizer (P1) mounted on a high precision rotational mount was used to pre-select the state at *p*-linear polarization. The beam was then focused by a lens (L1, focal length = 20 cm) to a spot size of $w_0$= 300 µm ($z_0$ = 45 cm). The polarization state of light reflected from a 45°-90°-45° BK2 prism (refractive index n=1.516) was post-selected by another linear polarizer (P2). The exact orthogonal configuration of the pre and the post-selected state was located by rotating P2 to obtain the minimum intensity. Weak measurements were then performed by rotating the polarization axis of P2 to ± $\epsilon$ angle away from this position. The resulting beam shift was detected by a CCD camera (2048 × 1536 square pixels, pixel dimension 3.45 µm). The measurements were performed for angle of incidence $\theta_i$ = *45°*, for a fixed propagation distance z = 50 cm (z >$z_0$) and for varying small angle $\epsilon$ ($0.0017 rad < \epsilon < 0.035 rad$).

The conventional weak measurement limit for the specifics of our experimental system and parameters of the chosen Gaussian pointer is theoretically estimated using standard Fresnel reflection coefficients of air-glass interface (for $\theta_i$ =45°) to be ( $\epsilon_{min} = \frac{\delta_p}{w_0} \sim 0.015\ rad$). Within this limit ($\epsilon > \epsilon_{min}$), clear transverse shifts in the reflected beam's centroid can be observed between the post selected states at ±$\epsilon$ (shown for two different $\epsilon$ values in **Fig. 1b**). As expected, the Gaussian shape of the pointer is retained and the shift is larger for smaller $\epsilon$ (~ 0.018 rad, close to the limit) as compared to the larger $\epsilon$ (~ 0.035 rad, well within the limit). For values of $\epsilon$ below the weak measurement limit ($\epsilon < 0.015\ rad$), the pointer profile deviates from Gaussian shape accompanied with a pronounced 'dip' in the spatial intensity profile (inset of **Fig. 1c**). In this regime, the shift of the beam centroid does not yield the expected amplification with decreasing $\epsilon$ (**Fig. 1c**). For the regime of $\epsilon$ lying within the weak measurement limit ($\epsilon > 0.015\ rad$), on the other hand, the weak values extracted from the centroid shifts (using Eq. 3) (**Fig. 1d**) show excellent agreement with the corresponding theoretical predictions (generated using Eq. 2).

**Figure 2** demonstrates extraction of large weak values outside the conventional limit (for Gaussian pointer) of weak measurement ($\epsilon < \epsilon_{min}$) using the real part of the complex zero of the response function. As $\epsilon$ is reduced far below this limit, prominent intensity minimum is observed in the spatial profile of the pointer (**Fig. 2a** shown for $\epsilon = \pm\ 0.001\ rad$), which corresponds to the real part of the complex zero of the response function. As $\epsilon$ is reduced further to zero, the intensity profile becomes double humped Gaussian yielding a real zero of the response function at $y_0 = 0$ (**Fig. 2a**). Systematic shift of the minima of the pointer intensity profile can also be observed in this regime of $\epsilon$ ($0 < \epsilon < \epsilon_{min}$) (**Fig. 2b**). In agreement with the prediction of Eq. 6b, the position of the intensity minimum scales as $\sim \frac{1}{cot\epsilon}$ ($\sim \epsilon$ in the limit of small $\epsilon$) (inset of **Fig. 2c**). The resulting weak values extracted from the position of the intensity minimum (using Eq. 6b) accordingly show excellent agreement with the corresponding theoretical estimates (**Fig. 2c**). These values are extremely large and lie far beyond the familiar weak value amplification limit for Gaussian pointer. These results provide conclusive evidence of the relationship between the real part of the complex zero of the response function and the weak value and demonstrate that using this relationship one can obtain arbitrarily large weak value amplification of small system parameters (tiny Spin Hall shift here) going beyond the conventional weak measurement limit.

In order to experimentally verify the relationship between the weak value, geometric phase gradient and the imaginary part of the complex zero of the response function, we adopted weak measurement on Spin Hall shift in one arm of a Mach-Zehnder interferometer (**Fig. 3a**). Weak measurements were performed in a similar manner by using similar optical arrangement as in **Fig. 1a** ($w_0 = 300$ μm, $z_0 = 45$ cm, $\theta_i = 45^o$). However, as apparent from Eq. 7, in order to probe the PB geometric phase one has to be sufficiently close to the weak interaction plane ($z << z_0$). The CCD camera used to record the interference pattern was therefore placed at a distance $z = 5$ cm. In agreement with the predictions of Eq. 7, transverse spatial ($y$) gradient of PB geometric phase is manifested as opposite tilt of the fringe patterns for post selected states at $\pm\epsilon$ ($\epsilon = 0.004\,rad$ shown in **Fig. 3b** top panel). The geometric phase gradient extracted from the tilt of the experimental fringe pattern (0.167 rad/μm) is in excellent agreement with the corresponding estimate for the simulated results (0.197 rad/μm) using parameters identical to the experimental situation (**Fig. 3b** bottom panel). The extracted PB phase gradient is subsequently used to determine the weak value of Spin Hall shift using Eq. (7). Once again, the weak value estimated from the interferometric experiment ($A_W^{SH} = 2.1 \times 10^3$) agrees well with the corresponding theoretical estimate (using Eq. 2) ($A_W^{SH} = 2.5 \times 10^3$). Note that here also, the weak value is estimated for the regime of $\epsilon$ lying outside the conventional weak measurement limit ($\epsilon < \epsilon_{min}$, $\epsilon_{min} = 0.015\,rad$) for Gaussian pointer. These results establish the relationship between the weak value, geometric phase gradient and the imaginary part of the complex zero of the response function making it possible to experimentally extract weak values from the geometric phase gradient using interferometric measurements.

In summary, we have presented experimental demonstration of a fundamental relationship between the weak value of an observable and complex zero of the response function of a system by employing weak value amplification of spin Hall shift of light beam. The relationship between the weak value and the real part of the complex zero of the response function offered a remarkably simple approach of obtaining extremely large weak value amplification beyond the conventional limit, which is  demonstrated using the minima of the Gaussian pointer intensity profile. The imaginary part of the complex zero of the response function is shown to be related to the gradient of geometric phase that evolves during weak measurements. This also offered an interferometric approach of obtaining large weak value amplification through quantification of geometric phase gradient. The universal relationship between the weak values and the complex zeros of the response function may bear interesting consequences in the general scenario of weak measurements.  The ability to obtain extremely large weak value amplification and to extract system parameters outside the region of validity of conventional weak measurements may open up a new paradigm of weak measurement potentially enhancing metrological applications of weak measurements for high precision measurement of small physical quantities or for enhanced probing of interactions.

**Figure Captions**

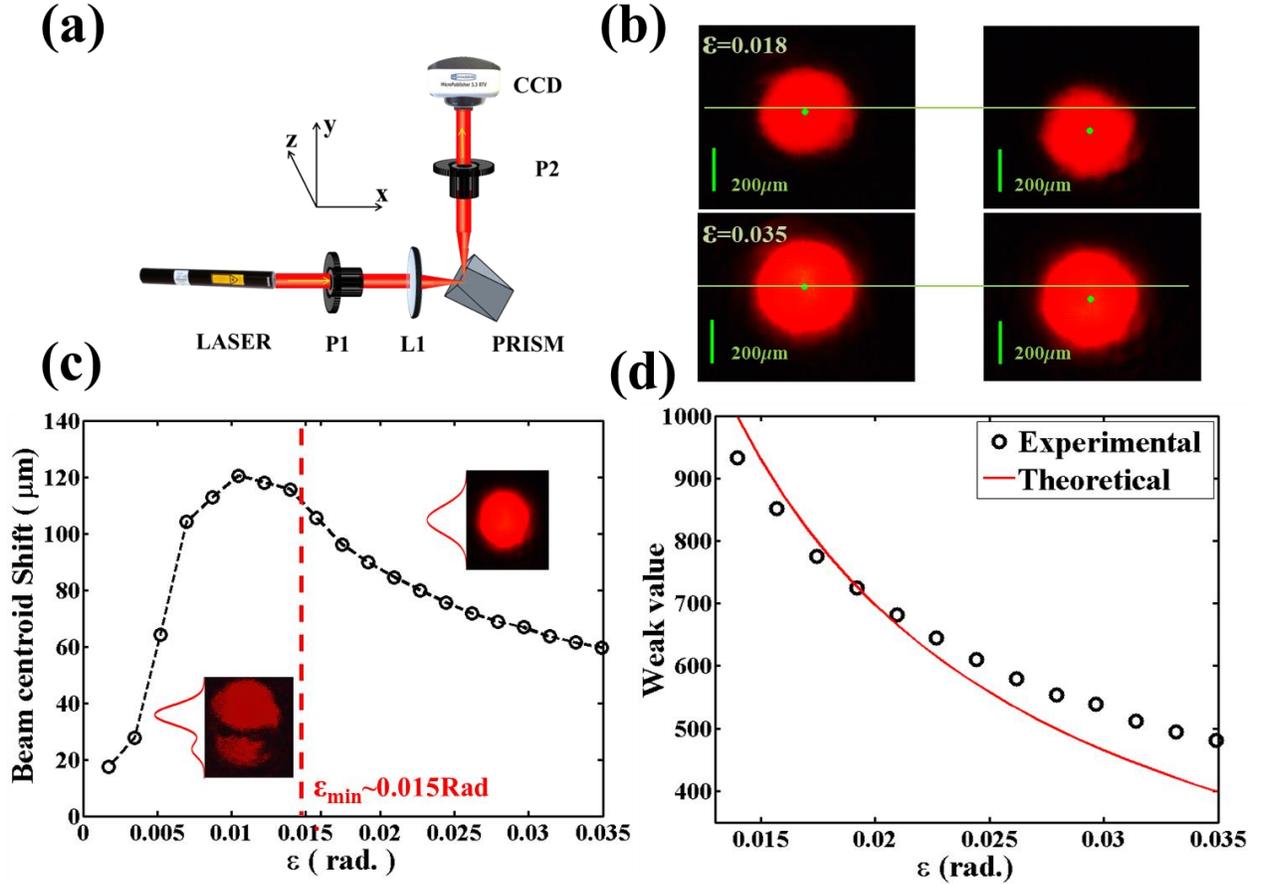

**Figure 1:** (a) A schematic of the set-up for weak value amplification of Spin Hall shift of light beam undergoing partial reflection at air-glass interface. (P1, P2): rotatable linear polarizers, L: lens. The prism acts as the weak measuring device. The experimental parameters are: $w_0 = 300$ μm, $z = 50$ cm, $\theta_i = 45^o$. (b) Transverse (along $y$) shift in beam's centroid between two post selected states $+\epsilon$ (left panel) and $-\epsilon$ (right panel) away from the orthogonal state for the pre-selected state $|\psi_{pre}\rangle = [1, \ 0]^T$, shown for two values of $\epsilon$, $\epsilon = 0.018 \ rad$ (top panel) and $\epsilon = 0.035 \ rad$ (bottom panel). (c) The shift in the beam's centroid with varying $\epsilon$ (open circles). The shape of the reflected beam at two different regimes, within ($\epsilon > \epsilon_{min}$) and outside ($\epsilon < \epsilon_{min}$) the domain of validity of conventional weak measurements is displayed. (d) The experimental weak values extracted from the centroid shifts (open circles) and the corresponding theoretical predictions (solid line, Eq. 2) as a function of varying $\epsilon$ within the weak measurement limit ($\epsilon > 0.015 \ rad$).

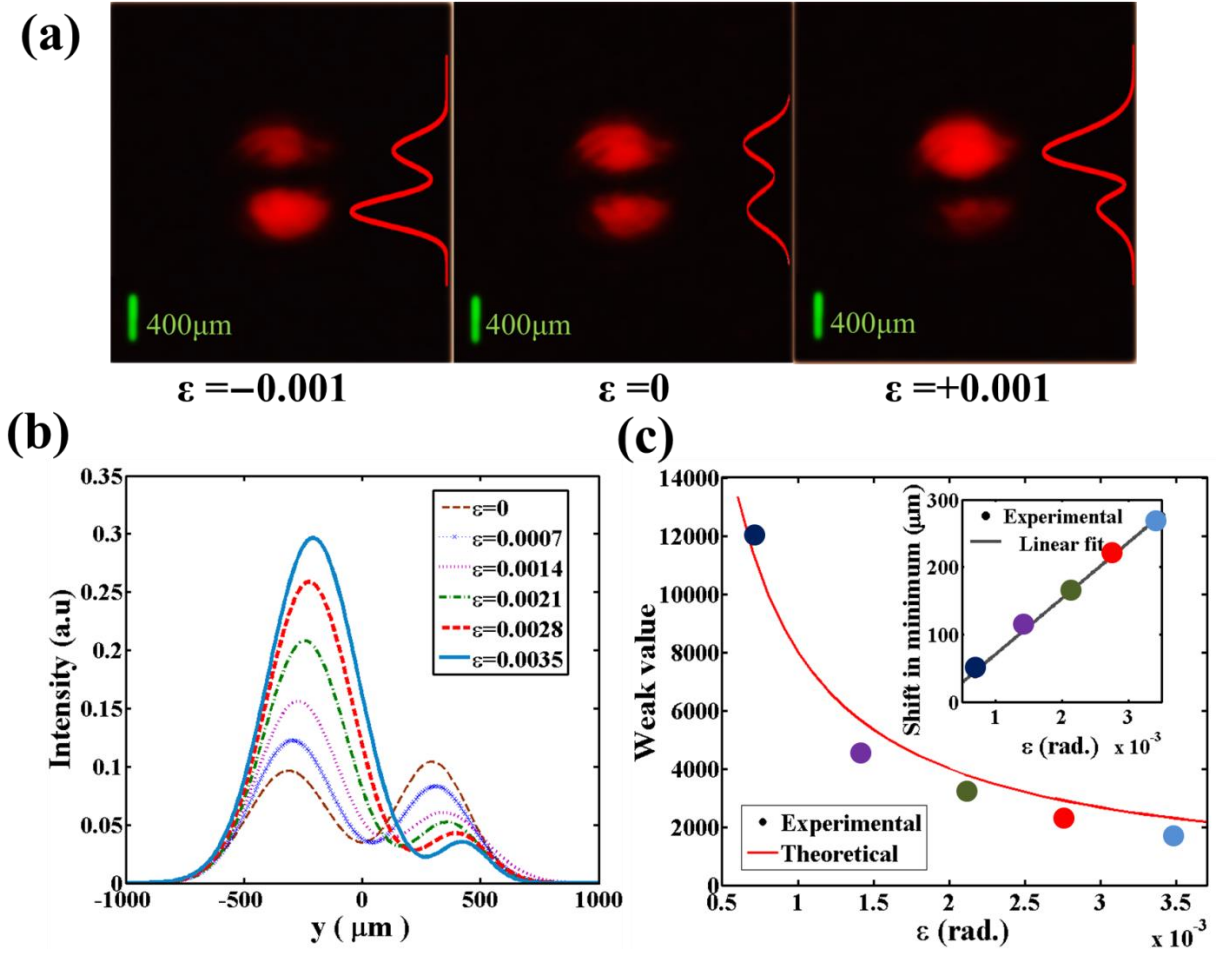

**Figure 2: Experimental demonstration of weak value extraction beyond the conventional limit of weak measurement ($\epsilon < \epsilon_{min}$) using the real part of the complex zero of the response function.** (a) Observation of intensity minima in the spatial profile of the pointer beam for post selections at $\epsilon < \epsilon_{min}$ ($\epsilon = \pm\, 0.001\ rad$ and $\epsilon = 0\ rad$). (b) Systematic shift of the minima of the pointer intensity profile along transverse $y$ direction with varying $\epsilon$. (c) The experimental weak values extracted (using Eq. 6b) from the position of the intensity minimum (solid circles) and the corresponding theoretical predictions (solid line) with varying $\epsilon$. The inset shows the $\sim\epsilon$ scaling of the spatial position of the intensity minimum.

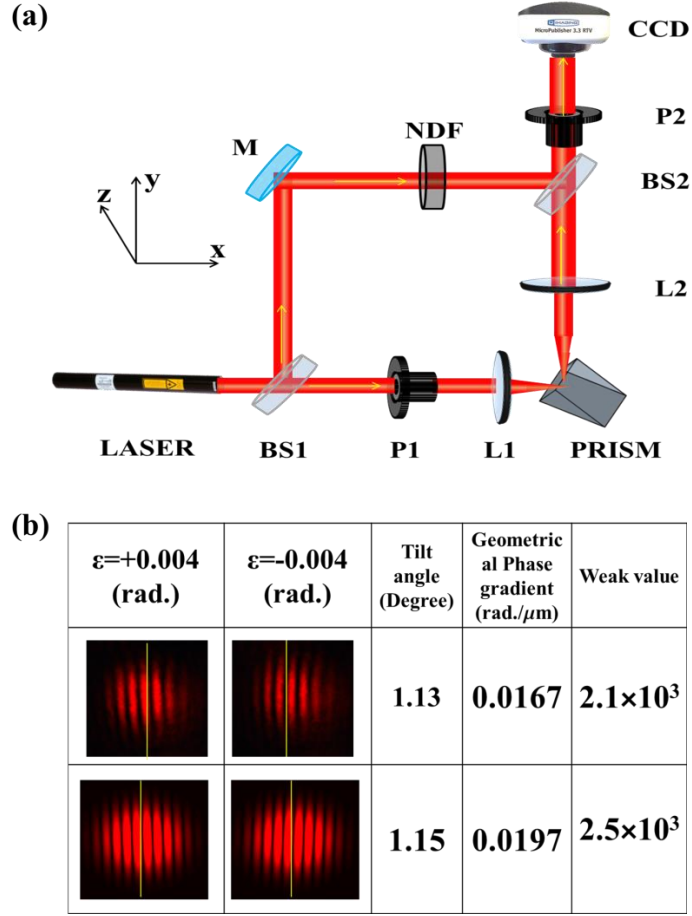

**Figure 3: Experimental verification of the relationship between the weak value, geometric phase gradient and the imaginary part of the complex zero of the response function.** (a) A schematic of the set-up employing weak measurement on Spin Hall shift in one arm of a Mach-Zehnder interferometer. (P1, P2): pre and post selecting linear polarizers, (L1, L2): lenses, (BS1, BS2): beam splitters, M: mirror, NDF: variable neutral density filter. The experimental parameters are: $w_0$= 300 μm, $z$ = 5 cm, $\theta_i$ = 45°. (b) The transverse spatial ($y$) gradient of geometric phase manifested as opposite tilt of the fringe patterns for typical post selected states at $\pm\epsilon$ (shown for a typical $\epsilon = 0.004\ rad$ in top panel). The corresponding theoretical simulations are shown in bottom panel. The estimates for tilt angle, geometric phase gradient and the weak value $A_W^{SH}$ (using Eq. 7) are noted.